\documentclass[useAMS,usenatbib]{mn2e}
\usepackage{epsfig}
\usepackage{amsmath}
\usepackage{amssymb}

\def\k2{\Bigl(\frac{2\pi}{\lambda}\Bigr)^2}
\def\rk2{\Bigl(\frac{\lambda}{2\pi}\Bigr)^2}

\newcommand{\sinc}{{\rm sinc}}

\title{Differential image motion in the short exposure regime}

\author[V.~Kornilov, B.~Safonov]{V.~Kornilov, B.~Safonov\thanks{E-mail: safonov@sai.msu.ru}\\
Sternberg Astronomical Institute, Universitetsky prosp. 13, 119992 Moscow, Russia}

\begin{document}
\date{Accepted --- Received ---}
\pagerange{\pageref{firstpage}--\pageref{lastpage}}
\pubyear{2011}
\maketitle
\label{firstpage}

\begin{abstract}
Whole atmosphere seeing $\beta_0$ is the most important parameter in site testing measurements. Estimation of the seeing from a variance of differential image motion is always biased by a non-zero DIMM exposure, which results in a wind smoothing. In the paper, the wind effects are studied within short exposure approximation, i.e. when the wind shifts turbulence during exposure by distance lesser than device aperture. 
The method of correction for this effect on the base of image motion correlation between adjacent frames is proposed. It is shown that the correlation can be used for estimation of the mean wind speed $\bar V_2$ and atmospheric coherence time $\tau_0$. Total power of longitudinal and transverse image motion is suggested for elimination of dependence on the wind direction. Obtained theoretical results were tested on the data obtained on Mount Shatdjatmaz in 2007--2010 with MASS/DIMM device and good agreement was found.
\end{abstract}


\section{Introduction}
\label{introduction}

The measurement of differential image motion by using Differential Image Motion Monitor (DIMM) is widely spread method for characterization of the atmospheric {\it optical turbulence} (OT) \citep{dimm,iac-dimm}. The differential image motion corresponds to the fluctuations of the difference of wavefront tilts (or difference of angles-of-arrival) in two close apertures. The DIMM is simple and robust instrument which suits well for long-term and field campaigns. Interpretation of its output data based on the theory of light propagation through turbulent media is quite straightforward: variance of the differential image motion is unambiguously connected with seeing $\beta_0$  \citep{dimm,Martin1987}.

However, there are several effects introducing biases and random errors in DIMM results \citep{Toko2002,KT2007}. One of them arises due to the fact that CCD camera captures star images which are used for centroids estimation with non-zero exposure. During typical exposure time $\tau \sim 10$~ms, a wind shifts the turbulence by distance comparable or more than aperture $D$. Therefore estimated centroids inevitably undergo the wind smoothing.

Necessity of reduction for this effect led to appearance of both theoretical  \citep{Martin1987} and experimental \citep{Soules1996} studies of impact of the exposure on power of the differential image motion measured with DIMM. The methods of correction developed in these works require an a priori information about wind or some temporal characteristics of the image motion estimated during the measurement process. Usually DIMM data is corrected for this bias using interlaced exposures. This method consists of alternation of single and double exposures (for instance 5 and 10~ms). Seeings are measured separately for these two sets and final estimation calculated as certain combination of them \citep{Sarazin1997,Toko2002}.

Realization of the measurements with modern CCD cameras changed situation significantly. It became possible to use shorter exposures and faster frame rates. 
On the other hand, the method of the interlaced exposures can be hardly used with these cameras.

Alternative solution of the problem consists in a measurement of covariance of the image motion between adjacent frames, which is defined by characteristic timescales of the process. In this paper we describe this method and prove theoretically its applicability. Also, it is shown that the covariance can be used to estimate some effective wind in the atmosphere.

In two first sections, we consider general expressions defining power of differential image motion and effect of wind smoothing for a single turbulent layer. The dependencies of the power (variance of differential image motion) on wind speed and wind direction and instrumental parameters are studied as well.

In the section~\ref{sec:short_approx}, we show that in {\it short exposure regime}, when the wind shear $v \tau \ll D$, where $v$ is the wind speed, the wind effects can be described as a quadratic functions of the shear. This approximation allows us to establish simple relation between the wind effects on variance and covariance.

The section~\ref{sec:whole_atm} generalizes obtained results for the case of the whole atmosphere. Expressions of the wind effects on variance and covariance depending on some effective wind are given. Then we describe the correction of measured variances to zero exposure and define an applicability and accuracy the method. Obtained formulae permit to evaluate square averaged wind $\bar V_2$ weighted with turbulence intensity and calculate the atmospheric coherence time $\tau_0$. This is demonstrated in the Section~\ref{sec:MASSwind} on the base of analysis of real data.

The last section contains discussion of the results and recommendations on optimal measurements and processing. Details of evaluation of needed integrals over two-dimensional spatial frequency are given in appendixes.

The numerical estimates are computed for DIMM device which we use for the OT measurement on Mount Shatdjatmaz \citep{kgo2010}. Main parameters of the instrument are aperture diameter $D = 0.09$~m, dimensionless baseline $b = 2.18$, exposure $\tau = 4$~ms before 2009 December 13 and $2.5$~ms after, period between exposures $T = 5$~ms.

\section{Basic relations for single turbulent layer}
\label{sec:theory}
\subsection{Spatial spectrum of the differential image motion}
\label{sec:space}

Equations for spatial spectrum of the differential image motion were given in classical works by \citet{Fried1965} and \citet{Martin1987}. In the paper of \citet{Martin1987}, these equations were developed by successive filtering of initial spatial spectrum of light wave phase distortions on the assumption of Kolmogorov OT model. It is known that spectral power density $\Phi$ of phase fluctuations is proportional to the intensity $\Delta J = C_n^2\,\Delta h$ in homogeneous and isotropic turbulent layer:
\begin{equation}
\Phi(f) = 0.0229\,r_0^{-5/3} f^{-11/3} = 0.0097\k2\!\Delta J\,f^{-11/3},
\label{eq:KolmPSD}
\end{equation}
where $f$ is the modulus of the 2D spatial frequency, $\lambda$ is the wavelength and $r_0$ is the Fried parameter. For practical purposes $\Delta J$ is more appropriate because this value is layer-by-layer additive and used for characterization of vertical OT profile. The phase spectrum is axisymmetric, i.e. depends only on the $f$ and has the dimension $\mathrm{m}^{-2}$. Eq.~(\ref{eq:KolmPSD}) is valid on condition that propagation effects of disturbed wavefront can be neglected (near-field approximation).

Hereinafter we use coordinate system different from one used by \citet{Martin1987}. The $x$ axis is placed along line connecting centres of apertures. In frequency domain, coordinates of a point are defined by modulus $f$ and position angle $\phi$ being reckoned from the $x$ axis. As usual two components of the differential image motion are considered: longitudinal is measured along the $x$ axis (hereafter {\it l-motion}) and transverse is measured along $y$ axis ({\it t-motion}).

The spectrum $F$ of the image motion can be represented as a product of phase spectrum, gradient filter, aperture filter and differential filter  \citep{Martin1987,Toko2002}. Formulae for these filters were transformed to used coordinate system. Gradient filter $G$ for $l$- and $t$-motions has the form
\begin{gather}
G_l(f,\phi) = \rk2\!(2\pi f)^2 \cos^2\phi,\notag\\ G_t(f,\phi) = \rk2\!(2\pi f)^2 \sin^2\phi.
\label{eq:grad}
\end{gather}
where factor $(\lambda/2\pi)^2$ is needed to pass from the slope in ${\rm m}^{-1}$ to the angle-of-arrival in radians, measured in the method.

Expression for the aperture filter $A$ depends on how image centroids are evaluated. This issue is quite important because the wavefront cannot be considered flat within aperture of the diameter $D$ and corresponding image in focal plane is neither diffractive nor even axisymmetrical. Normally two models are considered: gravity centre of the image \citep[g-tilt, see][]{Martin1987} or position corresponding to the normal of plane approximating wavefront within aperture \citep[z-tilt, see][]{Fried1975}. The difference between these approaches was analyzed in the paper of \citet{Toko2002}. Both filters are axisymmetrical and can be expressed through Bessel functions $J_n$  as
\begin{gather}
A^g(Df) =  \Bigl(\frac{2J_1(\pi D f)}{\pi D f} \Bigr)^2\!, \notag\\A^z(Df) = \Bigl(\frac{8J_2(\pi D f)}{(\pi D f)^2}\Bigr)^2\!.
\end{gather}
We will use $A^g(Df)$ by default since an exact form of aperture filter is not critical for analysis.

The differential filter corresponding to two identical apertures placed at the distance $B$ from each other along $x$ axis is given by the expression:
\begin{equation}
H(f,\phi) = 4\sin^2(\pi B f\cos\phi).
\end{equation}

The final spectral power density $F(f,\phi)$ of the differential image motion measured with DIMM is defined by the relation:
\begin{equation}
F_{l,t}(f,\phi) = \Phi(f)\,G_{l,t}(f,\phi)\,A(Df)\,H(f,\phi).
\end{equation}
It is worth noting that after substitution of (\ref{eq:KolmPSD}) and (\ref{eq:grad}), the spectral density will not depend on wavelength $\lambda$ of incoming light.

\subsection{Power of differential image motion}
\label{sec:variance0}

Variance of the differential image motion $\sigma_{l,t}^2$ can be computed as integral of the $F_{l,t}(f,\phi)$ over all spatial frequencies. Because integration is being carried out in polar coordinates, factor $f$ has to be added. Let us group the spectral filters in order to separate terms depending and not depending on polar angle $\phi$ for consequent averaging by this angle. Also, it is convenient to use the dimensionless frequency $q=fD$ and the dimensionless baseline $b=B/D$. In this case, the variance of the differential image motion is
\begin{equation}
\sigma_{l,t}^2 = 2.403\,\Delta J\,D^{-1/3} \int_{0}^{\infty}q^{-2/3}A(q)\Psi_{l,t}(q){\rm\,d}q,
\label{eq:vars}
\end{equation}
where the factor $\Psi_{l,t}(q)$ is a result of averaging of all terms depending on angle $\phi$ over this angle. In case of the $l$-motion, it becomes
\begin{multline}
\Psi_l(q) = \frac{2}{\pi}\int_{-\pi}^{+\pi} \sin^2(\pi b q\cos\phi)\,\cos^{2}\phi  {\rm\,d}\phi \\ = 1 - J_0(2\pi b q) + J_2(2\pi b q),
\label{eq:psi_l1}
\end{multline}
and for the $t$-motion:
\begin{multline}
\Psi_t(q) = \frac{2}{\pi}\int_{-\pi}^{+\pi} \sin^2(\pi b q\cos\phi)\,\sin^{2}\phi  {\rm\,d}\phi \\ = 1 - J_0(2\pi b q) - J_2(2\pi b q).
\label{eq:psi_t2}
\end{multline}
For derivation of $\Psi(q)$, the relations from Append.~\ref{sec:a1} were used. In cases when it is not important, we will discard subscripts designating component of the image motion.

Sum of the variances of both components $\sigma_c^2=\sigma_t^2+\sigma_l^2$ (hereafter {\it total differential image motion: $c$-motion}) is very convenient for theoretical analysis and interpretation of observational data. Evidently, its gradient filter $G_c(f) = (\lambda\,f)^2$ is axisymmetrical and corresponding function $\Psi_c(q)$ looks like
\begin{equation}
\Psi_c(q) = \frac{2}{\pi}\int_{-\pi}^{+\pi} \sin^2(\pi b q\cos\phi) {\rm\,d}\phi = 2 - 2\,J_0(2\pi b q).
\label{eq:psi_c1}
\end{equation}

Although it is not explicitly indicated, functions $\Psi(q)$ depend on the baseline $b$. Integration over the dimensionless frequency $q$ leads to formulae depending on this only parameter. Let us denote these integrals as $\mathcal K(b)$:
\begin{equation}
\mathcal K(b) = \int_{0}^{\infty}\,q^{-2/3}\,A(q)\, \Psi(q){\rm\,d}q.
\end{equation}
Integrals $\mathcal K(b)$ can be expressed analytically through hyper-geometric functions, but exact expressions can not be used due to its bad convergence. It is more practical to utilize various approximations \citep[see e.g.][]{Martin1987,dimm,Toko2002}. For the $c$-motion, $\mathcal K(b)$ is given by:
\begin{equation}
\mathcal K_c(b) \approx 3.367\,(1.4023 - b^{-1/3}),
\label{eq:kcb_g}
\end{equation}
Accuracy of this formula is reasonably good because hyper-geometric function (entering the precise expression) has a good convergence for $b > 1$ and deviates from $1$ by no more than $0.005$ in domain $1.2 < b < \infty$. Analogous expression for the case of z-tilt can be obtained as well:
\begin{equation}
\mathcal K_c(b) \approx 3.367\,(1.5024 - b^{-1/3}).
\label{eq:kcb_z}
\end{equation}

Formulae (\ref{eq:kcb_g}) and (\ref{eq:kcb_z}) coincide with the equations~(7) and (8) are given in  \citep{Toko2002}, except that our definition of $\mathcal K(b)$ differs from the instrument constant $K$ derived there in additional factor $6.95 = (0.0229\cdot2\pi)^{-1}$. Also, it must be taken into account that  $\mathcal K_c(b) = \mathcal K_t(b) + \mathcal K_l(b)$.

\section{Wind effect on the differential image motion}
\label{sec:theory_wind}

\subsection{Spectral filters for non-zero exposures}
\label{sec:sp_filters}

Effect of wind smoothing can be estimated by multiplying of temporal power spectrum of the image motion by the spectral filter corresponding to averaging during the exposure $\tau$. However, it is easier to use spatial averaging in the direction of the wind in rectangular window with length equal to the wind shear $v\tau$ and zero width as it was done by \citet{Martin1987}. In the used coordinate system, the wind blows at the angle $\theta$ to the $x$ axis. Introducing dimensionless wind shear $\omega = v\tau/D$, one can write the spatial averaging filter $S(q,\phi)$ as
\begin{equation}
S_1(q,\phi,\omega,\theta) = \sinc^2(\omega\,q\cos(\phi-\theta)),
\label{eq:s1filter}
\end{equation}
where $\sinc(x) = \sin(\pi x)/ \pi x$. It is clear that the role of this filter is to suppress high-frequencies in the spatial spectrum.

Covariance of image motion between adjacent exposures can be expressed through variance of binned data. In order to obtain this variance, it is not sufficient just to double exposure in formula (\ref{eq:s1filter}) because in actual measurements existing technical limitations lead to a certain pause between adjacent  exposures. Therefore one should use the filter, which corresponds to two separate averaging windows:
\begin{equation}
S_2(q,\phi,\omega,\hat\omega,\theta) = \sinc^2(\omega q\cos(\phi-\theta))\cos^2(\pi \hat\omega q\cos(\phi-\theta)),
\label{eq:s2filter}
\end{equation}
where the {\it period} of exposures $T \ge \tau$ and corresponding dimensionless wind shear $\hat\omega =  vT/D$. It can be seen that in the case of $T = \tau$ expression for $S_2(\omega,\hat\omega)$ becomes identical to $S_1(2\omega)$.

\begin{figure*}
\centering
\psfig{figure=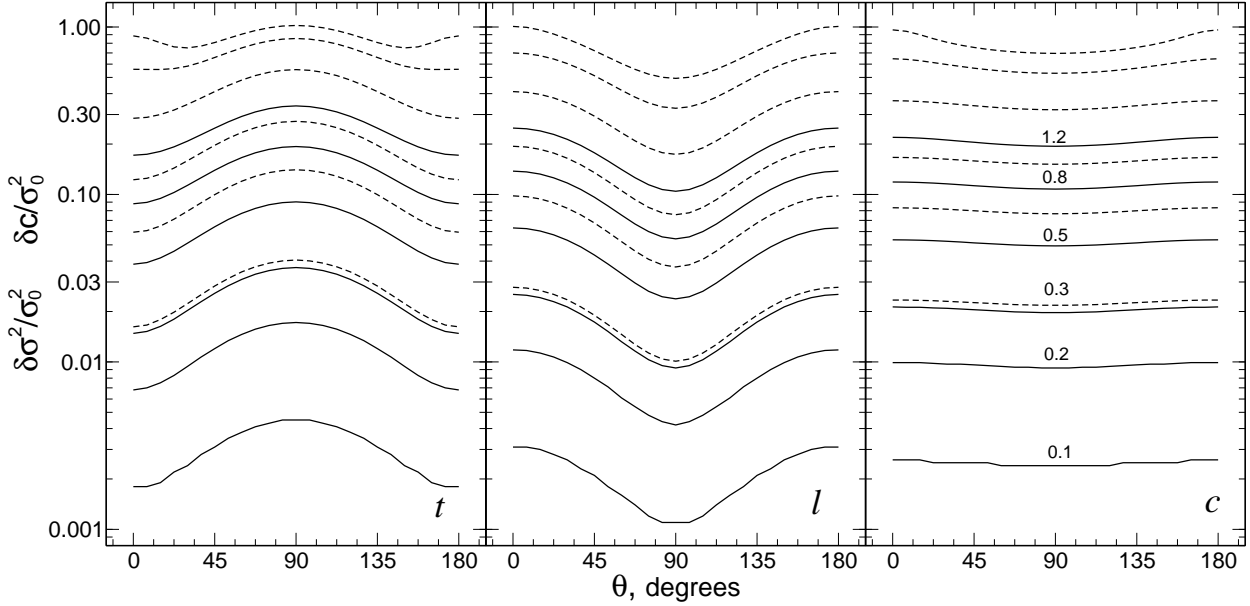,width=8.5cm,angle=-90}
\caption{Normalized wind effect $\delta\sigma^2(\omega,\theta)/\sigma_0^2$ (solid) and $\delta c(\omega,\hat\omega,\theta)/\sigma_0^2$ (dashed) for transverse ($t$), longitudinal ($l$) and total ($c$) differential image motions with different values of dimensionless wind shear $\omega$ and $\hat\omega = 1.25\,\omega$ depending on wind direction $\theta$. \label{fig:effect1}}
\end{figure*}

Multiplication of spectrum $F(q,\phi)$ by the filter $S$ and subsequent integration lead to the expression for the variance  $\tilde\sigma^2$ of the smoothed image motion. Extracting term for non-smoothed power $\sigma_0^2$, one can get
\begin{multline}
\tilde\sigma^2 = \iint F(q,\phi)\,S(q,\phi,\omega,\theta) {\rm\,d}q\,{\rm\,d}\phi \\=
\sigma_0^2 - \iint F(q,\phi)(1-S(q,\phi,\omega,\theta)) {\rm\,d}q\,{\rm\,d}\phi.
\label{eq:taut}
\end{multline}
The last integral represents the correction to zero exposure $\delta\sigma^2(\omega,\theta)$. Hereinafter we will call this value as wind effect on variance.

With help of equation for covariance between adjacent positional measurements $cov_1 = 2\sigma^2_2 - \sigma^2_1$, the corresponding filter $C_1$  can be obtained:
\begin{multline}
C_1(q,\phi,\omega,\hat\omega,\theta) \\= \sinc^2(\omega\,q\cos(\phi-\theta))\cos(2\pi \hat\omega\,q\cos(\phi-\theta)).
\label{eq:covfilter}
\end{multline}
Respectively, the wind effect on covariance:
\begin{equation}
\delta c(\omega,\hat\omega,\theta) = \iint F(q,\phi)(1-C_1(q,\phi,\omega,\hat\omega,\theta)) {\rm\,d}q\,{\rm\,d}\phi.
\label{eq:covcorr}
\end{equation}
The dependence of $\delta\sigma^2(\omega,\theta)$ and $\delta c(\omega,\hat\omega,\theta)$ on turbulence layer intensity can be removed by normalization by zero exposure variance $\sigma_0^2$.

Parts of the filters $S_1$ and $C_1$ depending on angle $\phi-\theta$ should be included in integrand in expression for function $\Psi(q)$ (see section \ref{sec:variance0}). Practically, this effect can be estimated by virtuosic analytics \citep[e.g. by series expansion, ][]{Conan2000} or numerical integration. The latter is more preferable for us since obtained results in any case are equally difficult for analysis due to many input parameters.

\subsection{Features of wind smoothing effect}
\label{sec:sp_features}

The dependencies of the normalized wind effects on wind direction and wind shear computed with formulae (\ref{eq:taut}) and  (\ref{eq:covcorr}) are shown in Fig.~\ref{fig:effect1}. Variations of the wind effects with wind direction can be clearly seen. The effects are at maximum when the direction of image motion coincides with the wind direction and at minimum when they are perpendicular. The maximum differs from the minimum almost twice. On the average, the normalized effects are larger for $t$-motion than for $l$-motion. The same is valid for the covariance.

The curves for different values of the wind shear are almost similar what indicates that dependencies on $\theta$ and $\omega$ can be separated. Moreover, when $\omega \lesssim 1$ the curves $\delta c/\sigma_0^2$ resemble $\delta\sigma^2/\sigma_0^2$. Dependence on the wind direction $\theta$ can be described by $\cos 2\theta$ with reasonably good accuracy. Note, that total variance  $\tilde\sigma^2_c$ of image motion virtually does not depend on the wind direction. Its maximal deviation from the mean is less than 1.5 per cent.

Since dependencies of variances $\tilde\sigma^2_t$ and $\tilde\sigma^2_l$ on wind direction are `opposite in phase', their ratio also changes markedly. Range of possible ratios $\tilde\sigma^2_t/\tilde\sigma^2_l$ increases with increase in wind shear and for $\omega=1$ reaches $1.25 \div 0.92$ of initial ratio $\sigma^2_t/\sigma^2_l$.

The $l$- and $t$-motions are independent, in other words their correlation (corresponding gradient filter is $(\lambda f)^2 \cos\phi\,\sin\phi$) equals to zero. However when wind smoothing is taking place, the measured $l$- and $t$-motions become partly correlated. Correlation coefficient changes as $~\sin 2\theta$, i.e. reaches its maximum when wind blows at angle $\pm 45^\circ$ to the baseline.

The same dependencies as in Fig.~\ref{fig:effect1} but as a functions of the wind shear $\omega$ are presented in Fig.~\ref{fig:effect2}. For small argument, the curves are parallel to each other what indicates their proportionality.  Weak dependence of the $\tilde\sigma^2_c$ on wind direction results in a quite small distance between upper and lower boundaries. Slope of the curves corresponds to quadratic dependence of the effect on $\omega$. This feature will be considered in details in following section.

\section{Short exposure approximation}
\label{sec:short_approx}

\subsection{Averaging over all wind directions}
\label{sec:theta-averaging}

\begin{figure}
\centering
\psfig{figure=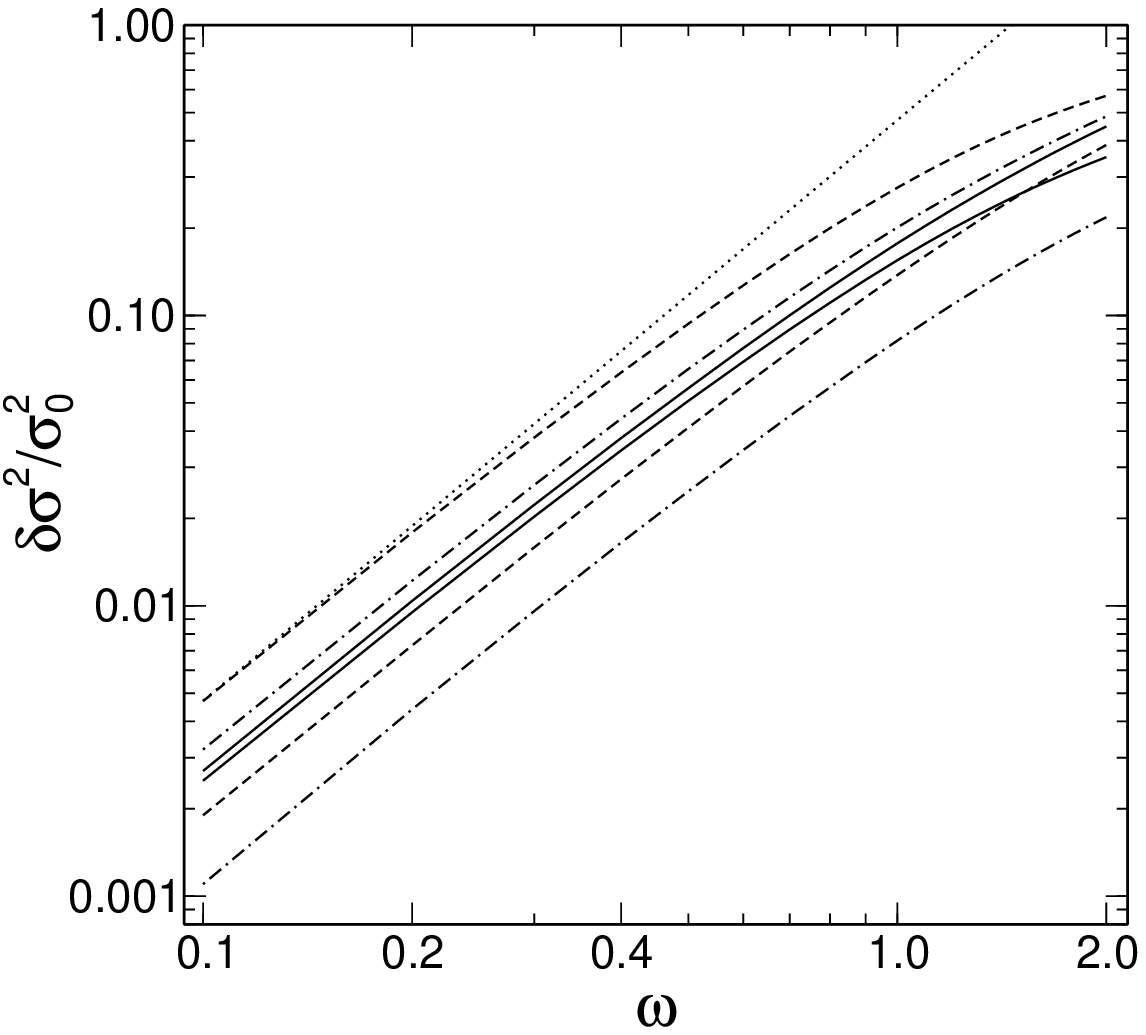,height=8.0cm,angle=-90} \\
\psfig{figure=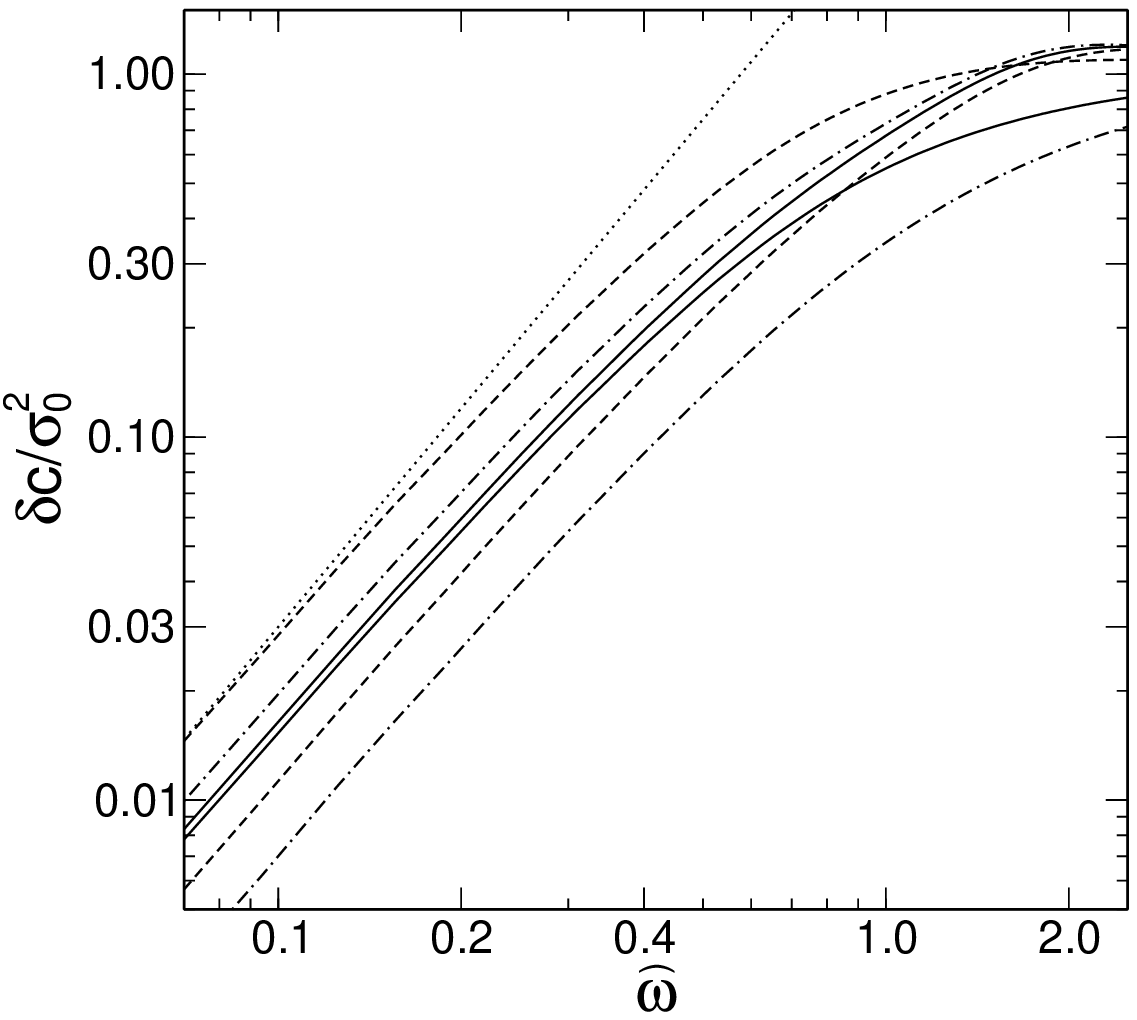,height=8.0cm,angle=-90}
\caption{Normalized wind effect $\delta\sigma^2(\omega,\theta)/\sigma_0^2$ (top) and $\delta c(\omega,\hat\omega,\theta)/\sigma_0^2$ (bottom) as a function of wind shear $\omega$ and $\hat\omega$. For bottom plots $\omega = 0.8\,\hat\omega$. Upper and lower boundaries for total (solid lines), transverse (dashed) and longitudinal (dash-dotted) differential motions are presented. The dotted lines depict a quadratic dependencies on argument.
\label{fig:effect2}}
\end{figure}

The mean wind effect can be estimated by averaging of the expression (\ref{eq:taut}) for the variance $\tilde\sigma^2$ over angle $\theta$. Such averaging affects only the filter $S_1$ and leads to the expression
\begin{equation}
\tilde\sigma^2 = \iint F(q,\phi)\,{\mathcal T_1(\omega q)}{\rm\,d}q\,{\rm\,d}\phi,
\end{equation}
where the function ${\mathcal T_1(\omega q)}$ is a spectral filter of the wind shear \citep{wind2010}. Further averaging over the angle $\phi$ results in a new function $\Psi^{\prime}(q) = \Psi(q)\,{\mathcal T_1(\omega q)}$.

The filter of the wind shear has the asymptotic ${\mathcal T_1(\omega q)} \approx 1 - \pi^2 \omega^2 q^2/6$ near the origin which is accurate while $\omega \ll 1/\pi$ because $q < 1$. Using this approximation, one can calculate the wind effect $\delta\sigma^2(\omega)$ or correction to zero exposure:
\begin{equation}
\delta\sigma^2(\omega) =  2.403\Delta J\,D^{-1/3}\frac{\pi^2 \omega^2}{6} \int_{0}^{\infty}q^{4/3}A(q) \Psi(q){\rm\,d}q.
\label{eq:corr1}
\end{equation}

Integral over dimensionless frequency in this formula
\begin{equation}
\mathcal L(b) = \int_{0}^{\infty}\,q^{4/3}\,A(q)\, \Psi(q){\rm\,d}q,
\label{eq:lint}
\end{equation}
differs from integral $\mathcal K(b)$ in the multiplicand $q^{4/3}$ only and also depends on a single parameter $b$. For the case of $c$-motion, it can be computed analytically and for $b = 2.18$ equals to $\mathcal L_c(b) = 0.367$, varying insignificantly in range between $0.374$ and $0.366$ for $1.2 < b < \infty$. As above, to designate the total differential image motion, subscript $c$ are used.

Let us consider the covariance when exposure $\tau = 0$. Then the filter $S_1 \equiv 1$ and averaging over all wind directions results in
\begin{equation}
\Psi^{\prime}(q) = \Psi(q)\,J_0(2\pi \hat\omega\,q).
\label{eq:dc_appr1}
\end{equation}

If the wind shear is finite but small, $\omega \ll 1$, an expansion of $\sinc^2$ near zero can be used and filter $C_1(q,\phi,\omega,\hat\omega,\theta)$ can be transformed in the following way:
\begin{multline}
C_1(q,\phi,\omega,\hat\omega,\theta) = \bigl(1 - \frac{1}{3}\pi^2 \omega^2\,q^2\cos^2(\phi-\theta)\bigr)\\
\times\cos(2\pi \hat\omega\,q\cos(\phi-\theta)),
\label{eq:C1_small}
\end{multline}
after averaging this filter over $\theta$:
\begin{multline}
\Psi^{\prime}(q) = \Psi(q)\\ \times\Bigl( J_0(2\pi \hat\omega\,q) -\frac{\pi^2 \omega^2\,q^2}{6}\,\bigl(J_0(2\pi \hat\omega\,q) - J_2(2\pi \hat\omega\,q)\bigr)\Bigr).
\end{multline}
Comparison of this expression with (\ref{eq:dc_appr1}) allows to conclude that wind effect on covariance depends mostly on the exposures period $T$. Even in the case of $\tau = T$, the value $\delta c$ changes only by $\approx 1/6$ with respect to the case $\tau = 0$.

Expansion of the Bessel function near zero permits to get the formula for $\delta c$:
\begin{equation}
\delta c = \Bigl(\frac{\pi^2 \omega^2}{6} + \pi^2\hat\omega^2\Bigr)\, 2.403\,\Delta J\,D^{-1/3} \mathcal L(b).
\end{equation}
From this expression and the formula (\ref{eq:corr1}),  universal relation between the wind effects on variance and covariance can be obtained:
\begin{equation}
\delta\sigma^2(\omega) = \delta c(\omega,\hat\omega)\frac{\omega^2}{\omega^2 + 6\,\hat\omega^2}.
\label{eq:depend}
\end{equation}

\subsection{Arbitrary wind direction}
\label{subsec:arbit_dir}

For arbitrary wind direction, the computing of the wind effects reduces to integration of the expression containing the differential filter, the gradient filter and the $S_1$ over angle $\phi$. Let us suppose that we can use quadratic expansion of $1-S_1$:
\begin{equation}
1-S_1 = \frac{1}{3}\pi^2 \omega^2\,q^2\cos^2(\phi-\theta).
\label{eq:S1_series}
\end{equation}
Then, the wind effect on variance is calculated by formula
\begin{equation}
\delta\sigma^2(\omega,\theta) =  2.403\,\Delta J\,D^{-1/3}\frac{\pi^2 \omega^2}{3}{\mathcal L^*}(b,\theta),
\label{eq:corr2}
\end{equation}
where the $\mathcal L^*$ is similar to the integral (\ref{eq:lint}) with functions $\Psi^*(q)$  instead $\Psi(q)$ (see Append.~\ref{sec:a1}) and depends not only on the $b$ but also on the wind direction $\theta$.

Values of these integrals ${\mathcal L}^*(b,\theta)$ are given in Table~\ref{tab:Lint} for all components of image motion and for different wind directions. Note, that according to (\ref{avet}), the integrals ${\mathcal L}^*$ averaged over all directions are half of the corresponding $\mathcal L$.

Formulae (\ref{eq:psic2}), (\ref{eq:psil2}) and (\ref{eq:psit2}) reflect dependence on the wind direction in the form $\cos 2\theta$ as depicted in Fig.~\ref{fig:effect1}. These formulae allow to deduce expressions for $\Psi^*(q,\theta)$ averaged over all wind directions as well:
\begin{equation}
\langle\Psi^*(q,\theta)\rangle = \frac{1}{2}\Psi(q).
\label{avet}
\end{equation}
Therefore, after averaging over $\theta$, formula (\ref{eq:corr2}) coincides entirely with equation (\ref{eq:corr1}).

\begin{table}
\caption{Integrals $\mathcal K$, $\mathcal L$, and ${\mathcal L}^*$ for the $b=2.18$. Integrals ${\mathcal L}^*$ for $\theta = \pi/4$ are equal to ones averaged over wind direction.
\label{tab:Lint}}
\centering
\begin{tabular}{p{2cm}p{1.5cm}p{1.5cm}p{1.5cm}}
\hline
Integral & $t$-motion & $l$-motion & $c$-motion \\[2pt]
\hline
            \multicolumn{4}{c}{g-tilt}\\
$\mathcal K$                   & 0.8141 & 1.3067 & 2.1208 \\
${\mathcal L}$,                & 0.1780 & 0.1890 & 0.3672 \\
${\mathcal L}^*$, $\theta = 0    $ & 0.0489 & 0.1402 & 0.1891 \\
${\mathcal L}^*$, $\theta = \pi/4$ & 0.0890 & 0.0945 & 0.1836 \\[2pt]
${\mathcal L}^*$, $\theta = \pi/2$ & 0.1292 & 0.0489 & 0.1781 \\
            \multicolumn{4}{c}{z-tilt}\\
$\mathcal K$                   & 0.9788 & 1.4804 & 2.4592 \\
${\mathcal L}$,                & 0.2206 & 0.2314 & 0.4518 \\
${\mathcal L}^*$, $\theta = 0    $ & 0.0597 & 0.1717 & 0.2314 \\
${\mathcal L}^*$, $\theta = \pi/4$ & 0.1103 & 0.1157 & 0.2259 \\[2pt]
${\mathcal L}^*$, $\theta = \pi/2$ & 0.1608 & 0.0597 & 0.2205 \\
\hline
\end{tabular}

\end{table}

Similar relation can be obtained for the wind effect on covariance by using of quadratic approximation of spectral filter $1-C_1$:
\begin{equation}
1-C_1 = \Bigl(\frac{\omega^2}{3} + 2\,\hat\omega^2\Bigr) \pi^2 q^2\cos^2(\phi-\theta).
\label{eq:C1_series}
\end{equation}
Clearly, the result for $\delta c(\omega,\hat\omega)$ differs from expression (\ref{eq:corr2}) only in factor $\omega^2/3 + 2\,\hat\omega^2$ instead of $\omega^2/3$. It means that for arbitrary wind direction, a relation between the wind effects on variance and covariance is described by formula (\ref{eq:depend}) as well.

\subsection{Applicability of the short exposure approximation}
\label{secShort}

\begin{figure}
\centering
\psfig{figure=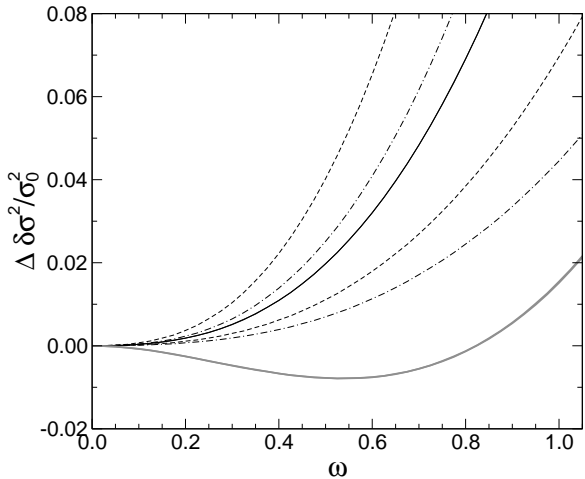,height=8.0cm,angle=-90}
\caption{Accuracy of quadratic correction of variance as a function of wind shear $\omega$. Upper and lower boundaries for total (solid lines), transverse (dashed) and longitudinal (dash-dotted) differential motions are shown. Grey lines correspond to the $c$-motion and coefficient reduced to 0.11 (see the text). Note that for $c$-motion, there are actually two lines for both coefficients but they are undistinguished in the picture scale.
\label{fig:corr1}}
\end{figure}

As it was shown above, the dependence of the wind effects on the wind shear demonstrates the quadratic behavior for short exposures and period between them. As a matter of fact, this is the definition of short exposure regime: the conditions for which the quadratic approximation is sufficient for practical usage.

In the case of measurement of stellar scintillation with Multi-Aperture Scintillation Sensor \citep{mass}, the effect of wind smoothing can be described by minimal set of parameters and a single criterion of applicability of short exposure approximation can be defined \citep{tau2011}. For differential image motion, the situation is more complex. Fig.~\ref{fig:effect2} indicates that for different components of the image motion and wind directions, criteria should be different.

Acceptable error of the approximation depends on problem at hand. As for correction to zero exposure, required absolute error of the normalized wind effect coincides with desirable relative accuracy $\epsilon$ of the result: $\Delta\delta\sigma^2/\sigma_0^2 = \epsilon$.

Deviations of the quadratic approximation of $\delta\sigma^2/\sigma_0^2$ from their exact functions are shown in Fig.~\ref{fig:corr1}. For the $l$- and $t$-motions, there are a large uncertainty in estimation of allowable wind shear. For instance, in order to provide accuracy $0.02$ of $\sigma_0^2$ in case of wind blowing perpendicularly to baseline, the wind shear should be less than $0.4$ for the $t$-motion and less than $0.8$ for the $l$-motion. This corresponds to wind speed $v = 9$ and $18 \rmn{\,m\,s^{-1}}$, respectively.

Far greater certainty can be reached when the $c$-motion is considered, because it demonstrates weak dependence on the wind direction. In this case one can argue that required accuracy is provided for $\omega < 0.5$ ($11\rmn{\,m\,s^{-1}}$) for any wind direction.

The quadratic approximation of the wind effect on covariance $\delta c/\sigma_0^2$ works worse, what is expectable from more complex behavior of exact dependence and much greater coefficient in corresponding formulae. Requirements for approximation accuracy will be formulated later in sections dealing with practical application of $\delta c/\sigma_0^2$.

Quadratic approximation coefficients were computed from formula (\ref{eq:corr2}) using equation (\ref{eq:vars}) for normalization and integrals from Table~\ref{tab:Lint}. They ensure correct behavior near $\omega = 0$ but do not provide minimal residual in desirable range of the wind shear. The range of approximation can be expanded by slight reduction of computed coefficient as is shown in the figure with the grey curve. For the $c$-motion, error of $\delta\sigma^2$ becomes less than $0.01\sigma_0^2$ in the range $0 \le \omega \le 1$ which corresponds to maximal wind speed $v = 22.5\rmn{\,m\,s^{-1}}$ for $4$~ms exposure and $36\rmn{\,m\,s^{-1}}$ for $2.5$~ms.

\section{Effect of the whole atmosphere}
\label{sec:whole_atm}

\subsection{Total wind smoothing effect}
\label{sec:while_atm}

In previous sections we considered smoothing of the differential image motion for infinitely thin turbulent layer with intensity $\Delta J = C_n^2(h)\,{\rm dh}$ and certain wind speed and direction. As was mentioned before (see Sect.~\ref{sec:space}), the effect of wave propagation was not taken into account, though for some cases its impact can be significant\footnote{http://curl.sai.msu.ru/mass/download/doc/dimm\_specs.pdf}. Weak perturbations approximation assumes independence of phase distortions generated by different turbulent layers. Because of this all second moments can be simply added layer-wise and for measurement with zero exposure:
\begin{equation}
\sigma_0^2 = 2.403\,D^{-1/3} {\mathcal K(b)}\int_{A}\,C_n^2(h){\rm\,d}h = 2.403\,D^{-1/3} {\mathcal K}\,J_\mathrm{tot},
\label{eq:vars_tot}
\end{equation}
where integration is performed over the whole atmosphere and $J_{tot}$ is the total turbulence intensity. This is the basic equation of the differential image motion method of measurement of the integral turbulence intensity in the atmosphere.

For non-zero exposure (or $\omega \neq 0$) expression is somewhat more complex because in general case it is necessary to take into account dependencies $\omega(h)$ and $\theta(h)$. Nevertheless, the smoothed variances $\tilde\sigma^2(h)$ and covariances can be integrated over the atmosphere. The same is valid for wind effects on the variance $\delta\sigma^2$ and covariance $\delta c$.

Within the short exposure approximation, total effect on the variance (for clarity, dependence on $b$ is omitted) is given by:
\begin{equation}
\delta\sigma^2 = 2.403\,D^{-1/3} \frac{\pi^2}{3} \int_{A}\omega(h)^2\,C_n^2(h)\,{\mathcal L}^*(\theta){\rm\,d}h.
\label{eq:corr_omega}
\end{equation}
In order to make integrand dependent only on atmospheric characteristics, we replace the dimensionless wind shear by its definition $\omega = v \tau /D$ and take all device parameters out of the integral. Additionally, let us take out value $\langle{\mathcal L}^*\rangle ={\mathcal L}/2$ averaged over all wind directions leaving under the integral ratio $2{\mathcal L}^*(\theta)/{\mathcal L}$ which is of the order of unity.

After these transformations, the integral represents second atmospheric moment of the wind speed distorted by angular dependence. Being normalized by total turbulence intensity $J_\mathrm{tot}$ it becomes some effective squared wind speed $\langle v^2\rangle$. Dividing both parts of expression (\ref{eq:corr_omega}) by $\sigma_0^2$, one can get
\begin{equation}
\frac{\delta\sigma^2}{\sigma_0^2} = \frac{\pi^2}{D^2}\frac{\tau^2}{6}\frac{\mathcal L}{{\mathcal K}} \langle v^2\rangle,
\label{eq:corr_wind2}
\end{equation}
where
\begin{equation}
\langle v^2\rangle = \frac{1}{J_\mathrm{tot}}\int_{A} C_n^2(h)\,v(h)^2\frac{{\mathcal L}^*(\theta(h))}{\mathcal L}{\rm\,d}h.
\label{eq:v2_def}
\end{equation}

Similarly, one can obtain the expression for normalized wind effect on covariance:
\begin{equation}
\frac{\delta c}{\sigma_0^2} = \frac{\pi^2}{D^2}\Bigl(\frac{\tau^2}{6}+T^2\Bigr)\frac{\mathcal L}{{\mathcal K}} \langle v^2\rangle.
\label{eq:cov_wind2}
\end{equation}

The dependence (\ref{eq:corr_wind2}) can be used for estimation of wind effect on variance, if vertical profiles of the OT and the wind speed or effective squared wind speed are known. On the contrary, formula (\ref{eq:cov_wind2}) is useful to estimate atmospheric parameter $\langle v^2\rangle$ from measured $\delta c$ value.


Since sensitivity of DIMM instrument $\mathcal K(b)$ increases slowly with $b$ and $\mathcal L(b)$ does not, it follows from equation (\ref{eq:corr_wind2}), that DIMM with the larger baseline is less sensitive to wind impact.

\subsection{Correction of image motion variance}
\label{sec:shortcorr}

Recall that in the papers of \citet{Martin1987,Soules1996}, the wind correction was estimated on the basis of model assumptions about vertical OT profile $C_n^2(h)$ and wind speed $v(h)$. Estimations of this kind are useful for planning or analysis but are not suitable for reduction of real data due to low accuracy.

The usage of the $\langle v^2\rangle$ seems useful for real data reduction. This value is similar to square averaged wind $\bar V_2$ which enters the definition of atmospheric coherence time $\tau_0$ and sometimes being measured by other methods together with DIMM measurements. However, without information on dominant wind direction these speeds may differ almost twice (see Table~\ref{tab:Lint}).

The method of interlaced exposures and the method of covariance represent versions of the same method which does not require additional data.

From formulae (\ref{eq:corr_wind2}) and (\ref{eq:cov_wind2}) it follows as well as for single turbulent layer (\ref{eq:depend}) that for short exposure regime,  the wind effects on variance and covariance are proportional to each other and constant of proportionality $\alpha$ depends only on parameters of image acquisition process $\tau$ and $T$:
\begin{equation}
\delta\sigma^2 = \delta c\frac{\tau^2}{\tau^2 + 6\,T^2} = \delta c \,\alpha.
\label{eq:depend_fin}
\end{equation}
It allows us to correct the measured variance to zero exposure using the value $\delta c$ also obtained from measurements.

Constant of the process $\alpha = \tau^2/(\tau^2 + 6\,T^2) < 1/7$ because $T \ge \tau$. Both parts of the equation (\ref{eq:depend_fin}) can be divided by measured $\tilde\sigma^2$. In this case procedure of correction looks like:
\begin{equation}
\sigma_0^2 = \tilde\sigma^2\,\left(1 +  \alpha\frac{\delta c}{\tilde\sigma^2}\right).
\end{equation}

Recalling definition $\delta c = \sigma_0^2 - cov_1$  (\ref{eq:covcorr}), one can rewrite previous expression in the following form:
\begin{equation}
\sigma_0^2 = \tilde\sigma^2\,\frac{1 - \alpha\,\tilde\rho}{1-\alpha}.
\label{eq:corr_fin}
\end{equation}
The right hand expression includes only measured values, where the $\tilde\rho$ is correlation coefficient of successive measured centroids. Pay attention that the $\tilde\rho = cov_1/\tilde\sigma^2$ differs from true correlation coefficient $cov_1/\sigma_0^2$ in normalization.

\begin{figure}
\centering
\psfig{figure=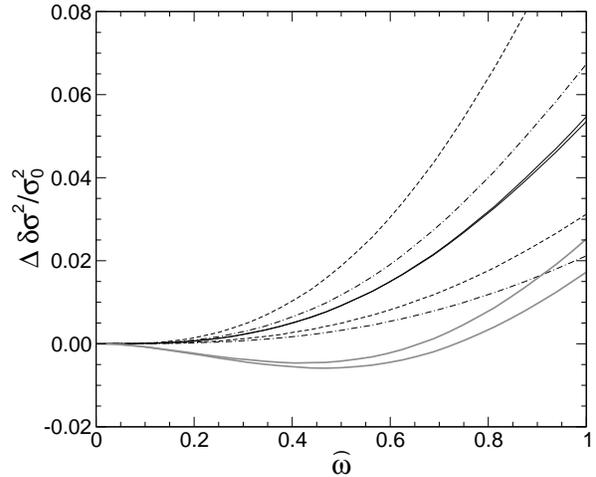,height=8.0cm,angle=-90}
\caption{Deviation of the exact dependence between normalized wind effects $\delta\sigma^2/\sigma_0^2$ and $\delta c/\sigma_0^2$ from (\ref{eq:depend_fin}) as a function of wind shear $\hat\omega$. Designations are the same as in Fig.~\ref{fig:corr1}. Grey lines correspond to $c$-motion with coefficient increased to 0.15 (see the text).
\label{fig:corrw}}
\end{figure}

Requirements for accuracy of the correlation coefficient are quite mild because $\alpha < 1/7$ (for our device $\alpha = 0.0964$). This is valid not only for measurement errors but also for errors arising from quadratic approximation of $\delta c(\hat\omega)/\sigma_0^2$. Thus, the error 20 per cent in correlation coefficient leads to error less than 2 per cent in $\sigma_0^2$.

In the method of interlaced exposures, one actually measures the variances $\tilde\sigma^2_{\tau}$ and $\tilde\sigma^2_{2\tau}$ affected by wind smoothing. The short exposure approximation allows to deduce simple relation between them:
\begin{equation}
\sigma_0^2 = \frac{4}{3}\tilde\sigma^2_{\tau} - \frac{1}{3}\tilde\sigma^2_{2\tau}.
\end{equation}

Deviation of the exact dependencies of $\delta\sigma^2/\sigma_0^2$ on  $\delta c/\sigma_0^2$ from linear functions with theoretical slope are shown in Fig.~\ref{fig:corrw}. It can be seen that differences are systematically positive and grow quite fast. For $\tilde\sigma_c^2/\sigma_0^2$ error $0.02$ is reached at $\hat\omega = 0.67$.

For shorter exposure 2.5~ms (instead of 4~ms), the error of the approximation significantly reduces (by factor of $\approx 2.5$) as well as wind smoothing effect itself. Maximal allowable value of the wind shear becomes $\hat\omega = 0.9$ ($16\rmn{\,m\,s^{-1}}$). Because of the constant $\alpha$ becomes $0.04$, the impact of $\delta c$ uncertainty also decreases.

Slope of the linear dependency predicted by theory, does not guarantee the minimal approximation errors in the needed range of wind shear. These errors can be significantly reduced by increase of slope of linear dependency. For example, the linear approximation of the dependence $\delta\sigma^2$ on $\delta c$ with slope $0.15$ provides error less than $0.01\,\sigma_0^2$ up to $\hat\omega = 0.8$ ($14.4\rmn{\,m\,s^{-1}}$). Note, that for our measurements, the most probable value $\hat\omega = 0.25$ and median value is $0.32$.

More complex approximation describing computed dependence $\delta\sigma^2(\tilde\rho)$ better, does not ensure additivity of the impacts of an individual layers and, in general case, may give worse results. Only linear combinations of variance and covariance provide independence on OT and wind speed vertical profile.

\subsection{Wind speed $\bar V_2$ evaluation}
\label{sec:wind_estim}

From the expression (\ref{eq:v2_def}) it follows that $\langle v^2\rangle$ differs from square averaged wind $\bar V_2$ in additional multiplicand $2{\mathcal L}^*(\theta)/{\mathcal L}$ which  varies by 2--3 times depending on wind direction in the case of $t$- and $l$-motions. At the same time, for the $c$-motion this factor deviates from 1 by no more than 3 per cent (see Table \ref{tab:Lint}) and angular dependence can be neglected. Therefore, in this case, one can assume $\bar V_2 = \langle v^2\rangle^{1/2}$ exactly.

The definition of $\delta c$ and the expression (\ref{eq:corr_fin}) set the connection between normalized correction for covariance and measured correlation coefficient $\tilde\rho$:
\begin{equation}
\frac{\delta c}{\sigma_0^2} = \frac{1-\tilde\rho}{1-\alpha},
\end{equation}
what allows to express square averaged wind speed $\bar V_2$ through instrumental parameters and measured $\tilde\rho$:
\begin{equation}
\bar V_2 = \frac{D}{\pi\,T}\left(\frac{\mathcal K}{\mathcal L}(1-\tilde\rho)\right)^{1/2}\!\!.
\label{eq:wind_res}
\end{equation}

\begin{figure}
\centering
\psfig{figure=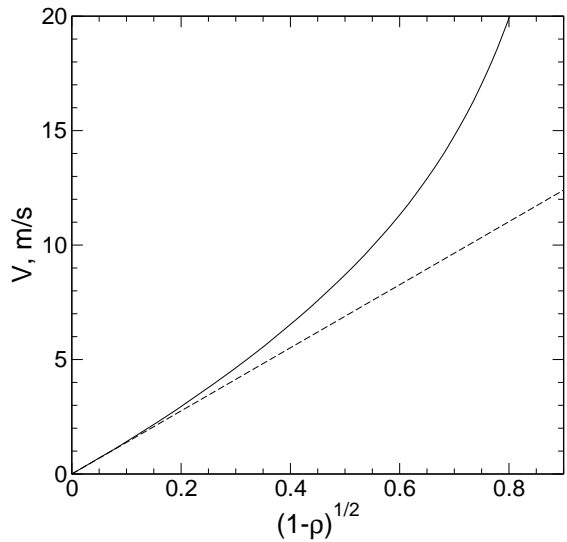,height=8.0cm,angle=-90}
\caption{Exact (solid line) and derived by (\ref{eq:wind_res}) (dashed line) dependencies of wind speed $\bar V_2$ on $(1-\tilde\rho)^{1/2}$.
\label{fig:wind}}
\end{figure}

This dependence gives reasonably good estimation of the $\bar V_2$ while quadratic approximation for $\delta c(\hat\omega)$ is valid. For greater wind speeds, rise of $\delta c(\hat\omega)$ slows down and evaluated $\bar V_2$ becomes underestimated. This effect is illustrated in Fig.~\ref{fig:wind}. For our DIMM device substitution of numerical values in formula (\ref{eq:wind_res}) gives $\bar V_2=13.8\,(1-\tilde\rho)^{1/2}$. Corresponding dependence is shown in figure by dashed line.

Comparison of the exact dependence with approximate shows that short exposure approximation underestimates wind speed by 13 per cent and 32 per cent for $5\rmn{\,m\,s^{-1}}$ and $10\rmn{\,m\,s^{-1}}$, respectively. Exposure time has almost no effect on appearance of this curve; on the contrary, change in the period $T$ leads to proportional change of $y$ axis scale.

Naturally, impact of fast turbulence layers will be significantly weakened. Therefore, the wind speed derived with DIMM can be considered only as lower estimate of the $\bar V_2$.
To do it more realistic, CCD camera frame period $T$ should be reduced twice at least. Then, the underestimate $\approx 30$ per cent will be provided for turbulent layer with $20\rmn{\,m\,s^{-1}}$ wind. Additionally, certain improvement can be achieved by increase of DIMM aperture $D$. Nevertheless, the method is able to give quite good estimates even with existent device, as will be shown in Sect.~\ref{sec:MASSwind} on real data.

Considered average wind speed $\bar V_2$ differs slightly from $\bar V_{5/3}$ entering the definition of atmospheric coherence time $\tau_0$. Nevertheless, investigation by \citet{TKel2007} has shown that $\bar V_2$ value may be used for estimation of $\tau_0$. Recall that method of derivation of $\tau_0$ implemented in FADE (Fast Defocusing of stellar image) \citep{FADE2008} and MASS instrument \citep{tau2011} are based on the $\bar V_2$ either.

\section{Experimental data}

\subsection{Measurements data and correction for finite exposure}
\label{sec:experiments}

Experimental proof of the basic relations was made using MASS/DIMM data obtained on Mount Shatdjatmaz in 2007--2010 \citep{kgo2010} for characterisation of OT.

The program {\sl dimm}\footnote{http://curl.sai.msu.ru/mass/download/doc/infrasoft\_eng.pdf} records measurements results in output file every 2~s and also its 1-minute averages. These results include variances $var$ and covariances of the differential image motions and coordinate noise $var^{*}$ in longitudinal and transverse directions. Value of coordinate noise is computed during processing of frames using formula (8) from \citep{MD2007} which gives slightly underestimated values for thresholding centroiding method since it does not take into account fluctuations of number of pixels exceeding a threshold.

Though in the real conditions, impact of the noise is very small (median of ratio $var^{*}/var \approx 0.0008$ and in 99.7 per cent of cases it is less than $0.01$), nevertheless, it is taken into account in further processing in obvious way: $\tilde\sigma^2 = var - var^{*}$. Covariance is not biased by the noise but incorrect account for the noise in $\tilde\sigma^2$ can lead to bias in the correlation coefficient.

Distributions of the correlation coefficients in 2007--2009 (containing $\approx 90\,000$ 1-minute points) have similar characteristics for all image motions: medians are equal to $0.85$, only for 4 per cent of events $\tilde\rho < 0.6$ and for 1 per cent --- $\tilde\rho < 0.44$. On the other hand, only for 5 per cent of the measurements, $\tilde\rho > 0.95$ and 0.5 per cent  $>0.97$. Distributions of 2-s points demonstrate similar behavior slightly different in details.

Prior to theoretical analysis described in this paper, during processing MASS/DIMM data in 2010, we used the following empirical law (based on numerical calculations, see Sect.~\ref{sec:sp_filters}):
\begin{equation}
\sigma_0^2 = \tilde\sigma^2\left(1+0.15(1-\tilde\rho)\right).
\label{eq:num_corr}
\end{equation}
To define the coefficient 0.15, the differential distribution of the $\tilde\rho$ as weights was used. Dependence (\ref{eq:num_corr}) corresponds to the formula (\ref{eq:corr_fin}) with $\alpha=0.13$.

\begin{figure}
\centering
\psfig{figure=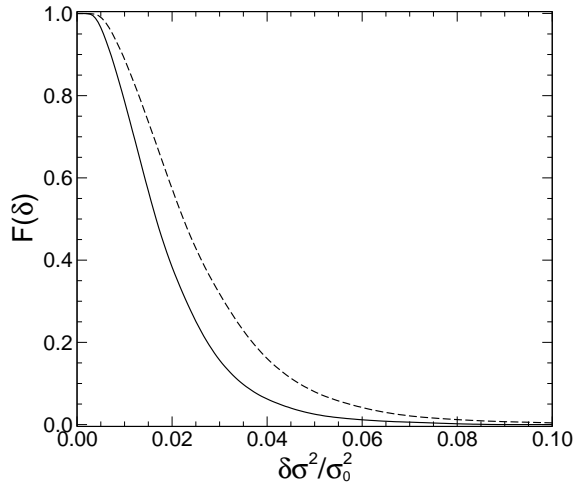,height=8.0cm,angle=-90}
\caption{Distribution of normalized corrections $\delta \sigma^2/\sigma_0^2$ computed using formula (\ref{eq:corr_fin}) with coefficients $\alpha=0.096$ (solid line) and $\alpha=0.13$ (dashed line).
\label{fig:distrib}}
\end{figure}

In Fig.~\ref{fig:distrib}, the distributions of the corrections $\delta \sigma^2/\sigma_0^2$ obtained from the $\tilde\rho$ using formula (\ref{eq:corr_fin}) with theoretical values $\alpha=0.096$ and $\alpha=0.13$ are shown. These distributions are quite close. Median of required correction is about 0.02. For 90 per cent of measurements, the correction is less than 0.04. It is possible to estimate that in marginal cases $\tilde\rho < 0.5$, the correction reaches 0.08. Note that these numbers are quite expectable taking into account that site under consideration has small median wind speed ($2.3\rmn{\,m\,s^{-1}}$).

\subsection{Dependence on wind direction}

Analysis of angles between wind direction and baseline for our measurements shows that their distribution is close to uniform. There is no any preferable angles, variations does not exceed $\pm 20$ per cent. For the detection of angular dependencies in the $l$- and the $t$-motion, the following combination of the wind effects are used:
\begin{equation}
\Theta = \frac{\delta c_t(\theta)-\delta c_l(\theta)}{\delta c_c(\theta)} \approx  \frac{\tilde\rho_l-\tilde\rho_t}{1-\tilde\rho_c}.
\end{equation}

Using results of Sect.~\ref{sec:while_atm}, it can be shown that
\begin{equation}
\Theta = \frac{\langle{{\mathcal L}^*_t(\theta)}-{{\mathcal L}^*_l(\theta)}\rangle}{\langle{{\mathcal L}^*_c(\theta)}\rangle},
\end{equation}
where averaging is done over the whole atmosphere with weight $v^2C_n^2$. Assuming that for majority of the measurements the ground layer is dominant, one can use surface wind direction as estimate of mean wind direction.

\begin{figure}
\centering
\psfig{figure=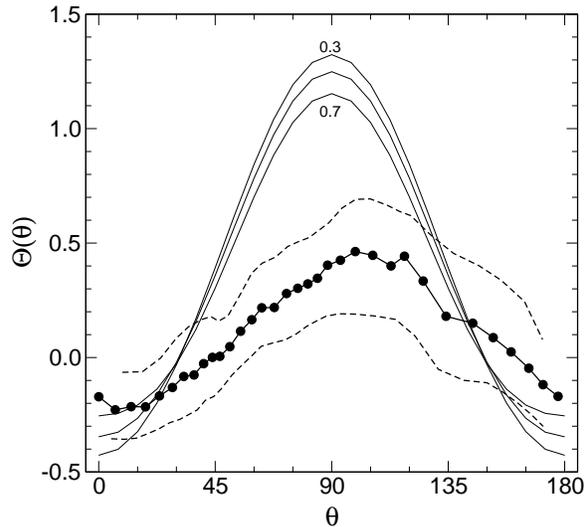,height=8.0cm,angle=-90}
\caption{Median $\tilde\Theta$ in groups of 500 points as a function of angle between wind direction and baseline (circles). The 1st and 3rd quartiles are indicated by dashed lines. Thin lines stand for expected behavior of $\Theta$ for different values of $\hat\omega$
\label{fig:angled1}}
\end{figure}

In Fig.~\ref{fig:angled1}, computed dependencies of $\Theta$ ratio on wind direction for different values of $\omega$ (they affect $\Theta$ only slightly) are shown. Medians, 1st and 3rd quartiles (in groups of 500 points) of $\tilde\Theta$ derived from measured $\tilde\rho$ are also presented. To minimize the  impact of measurements errors and uncertainties of the wind direction, data points were selected when surface wind was $v_\mathrm{GL} \ge 4\rmn{\,m\,s^{-1}}$.

Observed dependence looks similar to expected though its amplitude is lower. Difference in amplitude can be explained by the fact that wind direction can be different in different layers of atmosphere. Also amplitude could be reduced by small negative bias in $\tilde\rho_c$. We did not find dependence of the correlation coefficient of the $c$-motion on the wind direction.

\subsection{Comparison of wind speeds estimated from MASS and DIMM data}
\label{sec:MASSwind}
For verification of the estimate the effective wind, we have performed comparison of $\bar V_2$ estimates from data obtained simultaneously with DIMM and MASS. In order to do this, the average wind speed $\bar V_\mathrm{free} = 1.59\cdot10^{-9} J_\mathrm{free}^{-3/5} \tau_0^{-1}$ in free atmosphere was computed using $\tau_0$ and free atmosphere intensity $J_\mathrm{free}$ measured with MASS \citep{tau2011}. Then, the wind speed was corrected with ground layer intensity $J_\mathrm{GL}$ and surface wind speed $v_\mathrm{GL}$ from meteodata with help of expression $\bar V^2_\mathrm{MASS} = (\bar V^2_\mathrm{free} J^{}_\mathrm{free} + v^2_\mathrm{GL} J^{}_\mathrm{GL})/ J_\mathrm{tot}$.

The frame rate of our DIMM camera reaches 200 frames per second. The program {\sl dimm} output data format allows us to compute $\tilde\rho_c$ and then $\bar V_2$ using formula~(\ref{eq:wind_res}).

\begin{figure}
\centering
\psfig{figure=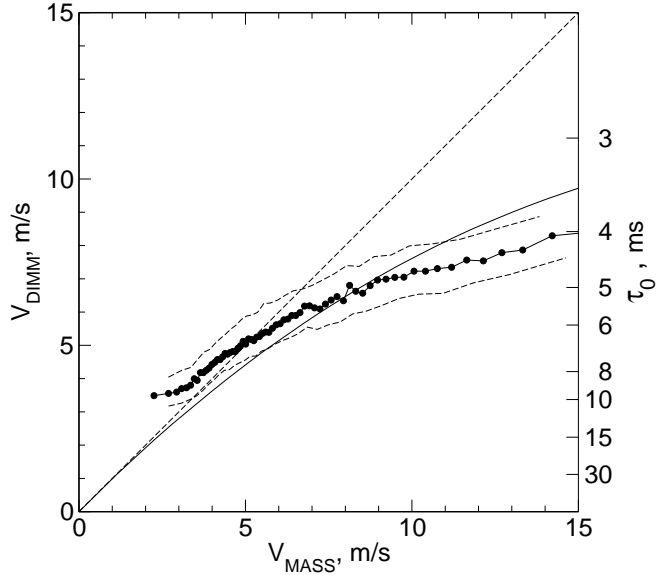,width=8.0cm,angle=-90}
\caption{Comparison of $\bar V_2$ estimated from DIMM and MASS data: dots stand for median wind speed in groups containing 500 measurements, dashed lines stand for 1st and 3rd  quartiles of distributions in these groups. Solid line stands for expected dependence considering errors of short exposure approximation.
\label{fig:wind2}}
\end{figure}

Results of comparison of 1-minute  $\bar V_2$ estimates 
and wind speeds $\bar V_\mathrm{MASS}$ is shown in Fig.~\ref{fig:wind2}, in the form of median and 1st and 3rd quartiles (in groups of $500$ points). The quartiles indicate that the dependence is quite narrow, mean spread is $\pm 1\rmn{\,m\,s^{-1}}$. Median curve is similar to expected exact dependence depicted by thin solid line but for higher wind speed it lies somewhat lower (this reflect the loss of impact of fast turbulence layers). For clarity, left vertical axis presents the atmospheric time computed assuming median $\beta_0 = 0.93$ arcsec.

Median values of the wind speeds $\bar V_\mathrm{MASS} = 5.84\rmn{\,m\,s^{-1}}$ and $\bar V_2 = 5.67\rmn{\,m\,s^{-1}}$ are very close, though there are reasons for believing that surface wind speed $v_\mathrm{GL}$ is underestimated due to anemometer was located at altitude 5~m. Nevertheless in median region wind speeds coincide quite well, in spite of absence of any empirical calibrations.

\section{Discussion and conclusions}

\label{sec:outputs}

Applicability of quadratic approximation of the wind effects for the measurement of the differential image motion depends on several instrumental parameters. Exposure time and period between exposures are most important of them.

Data analysis indicated that  for two brightest stars used in our measurements, 4~ms exposure lead to saturation in centres of images. Because of this we reduced exposure to 2.5~ms starting from December, 2009. Median noise impact increased insignificantly to $\approx 0.0011$ and for the faintest target stars with magnitude $\sim 3^m$ became $\approx 0.004$. Such negligible increase of the noise means that exposure can be reduced further.

The 2.5~ms exposure requires correction less by factor of approximately 2.5 ($\alpha=0.04$). For our measurements it means that in 90 per cent of time correction is less than 0.015 and only in rare cases $\sim 0.033$.

Reduction of the exposures period $T$ considerably expands capabilities of the DIMM. For the most CCD, maximal frame rate is defined by exposure and height of region of interest on detector.  Usually, the height of the window is set to 60--80 CCD rows. It is conditioned by amplitude of telescope tube vibrations occurring at wind speed greater than $\sim 7\rmn{\,m\,s^{-1}}$. More robust mount will allow to reduce the height of the window to 20--30 rows and increase frame rate up to 400 frames per second.

Such acquisition rate makes it possible to estimate mean squared wind $\bar V_2$ with reasonably good accuracy up to speed $\sim 10\rmn{\,m\,s^{-1}}$, what corresponds to $\tau_0 = 3$~ms for site with median seeing $\beta_0 = 1$ arcsec. Lower values of $\tau_0$ will be estimated with larger bias but these situations are less interesting for high-technology astronomical observations. Anyway, the possibility to estimate the atmospheric coherence time using DIMM is promising because DIMM is most common site testing instrument.

Let us briefly summarize main conclusions from analysis of the differential image motion in the short exposure approximation given in the paper.

\begin{enumerate}
\item Transverse image motion is more subject to the wind smoothing. The effect is at its maximum at wind blowing across the device baseline. On the contrary, for $l$-motion, the effect is maximal at wind blowing along the baseline.
\item Total differential image motion $\sigma^2_c = \sigma^2_l + \sigma^2_t$ has a simpler description. Its main advantage is a weak dependence (less than $\pm 0.02$) on the wind direction.
\item Effect of the wind smoothing for short exposures is proportional to exposure squared. It is possible to significantly diminish effect without loss of accuracy by using a shorter exposures.
\item Effect of the  wind smoothing for short exposures depends on device aperture diameter as $\propto D^{-2}$, what leads to significant increase in wind effect for DIMMs with small apertures.
\item Correction of measured variance of image motion can be performed in simple way using measured correlation coefficients between adjacent frames.
\item Value of the correlation coefficient for $c$-motion does not practically depends on wind directions and can be used for the estimation of the averaged wind $\bar V_2$ and the atmospheric coherence time for the whole atmosphere.
\end{enumerate}

Taking into account the fact that power of the $c$-motion $\sigma^2_c$ can be derived with greater accuracy than for $\sigma^2_l$ and $\sigma^2_t$ individually, we recommend to use it and corresponding relation (\ref{eq:kcb_g}) or (\ref{eq:kcb_z}) for calculation of the integral turbulence intensity. Widely spread estimation of the seeing as average between $\beta_0$ obtained from $l$- and $t$-motions in conditions of significant wind smoothing is biased and non-optimal.

Expressions for the wind effect $\delta c$ obtained in this work can be adopted to another pattern of readout from DIMM camera. For example, with full-frame CCD it is possible to obtain 2 (or more) successive exposures without pause between them and then to readout all subframes together during some time.

In the present work we did not consider influence of inner and outer scale of turbulence and possible departures from Kolmogorov law. We also did not take into account the effect of propagation of light wave which can be significant for high-altitude turbulence. Effect of propagation suppresses high frequencies of spatial spectrum of the image motions and, as a consequence, an influence of the wind smoothing become smaller. This reflects in decrease of integrals $\mathcal K$ and $\mathcal L$, and the latter changes far more significantly. Amount of effect is about 20 per cent for the turbulence in the tropopause. Nevertheless, the quadratic approximation (\ref{eq:corr_wind2}, \ref{eq:cov_wind2}) is still valid and can be used.

\section{Acknowledgements}
Authors are grateful to astroclimatic community for the interest to our research which reveal itself at two latest conferences and stimulated completion of this work. We also would like to thank our colleagues from Sternberg Astronomical Institute which provided us with observational data obtained with MASS/DIMM device on Mount Shatdjatmaz. B.S. acknowledges financial support from ``Dynasty'' foundation.

\bibliography{reference_list}
\bibliographystyle{mn2e}

\appendix
\section{$\Psi$ functions}
\label{sec:a1}
Averaging of components of differential motion spectrum depending on angle $\phi$ over this angle results in the expressions
\begin{multline}
\Psi_l(q) = \frac{2}{\pi}\int_{-\pi}^{+\pi} \sin^2(\pi b q\cos\phi)\,\cos^{2}\phi  {\rm\,d}\phi \\=
1 - J_0(2\pi b q) + J_2(2\pi b q),
\label{eq:psi_l}
\end{multline}
and
\begin{multline}
\Psi_t(q) = \frac{2}{\pi}\int_{-\pi}^{+\pi} \sin^2(\pi b q\cos\phi)\,\sin^{2}\phi  {\rm\,d}\phi \\=
1 - J_0(2\pi b q) - J_2(2\pi b q).
\label{eq:psi_t}
\end{multline}
Here the integrals $\mathcal J_0$ and $\mathcal J_1$ from Append.~\ref{sec:integrals} were used.

For $c$-motion $\sigma_c^2 = \sigma_l^2+\sigma_t^2$, result of the averaging is equal to sum $\Psi_l(q)+\Psi_t(q)$ or
\begin{equation}
\Psi_c(q) = \frac{2}{\pi}\int_{-\pi}^{+\pi} \sin^2(\pi b q\cos\phi) {\rm\,d}\phi = 2 - 2J_0(2\pi b q).
\label{eq:psi_c}
\end{equation}

In presence of wind smoothing, a similar functions $\Psi^*(q,\theta )$ can be obtained. In order to do this one should multiply integrands by $\cos^2(\phi-\theta)$ what follows from (\ref{eq:S1_series}) and (\ref{eq:C1_series}). This factor can be represented as
$ \cos^2(\phi-\theta) = \cos^2\theta - \cos 2\theta\sin^2\phi. $ The cross term with $\cos\phi\sin\phi$ is omitted because it gives zero after integration. Using formulae from Append.~\ref{sec:integrals} it is possible to write:
\begin{multline}
\Psi^*_c(q) = \cos^2\theta\,\Psi_l(q) + \sin^2\theta\,\Psi_t(q) \\= 1-J_0(2\pi b q)+\cos 2\theta\,J_2(2\pi b q).
\label{eq:psic2}
\end{multline}
For wind blowing across the DIMM base, function $\Psi^*_c(q)$ coincides with $\Psi_l(q)$ and for wind along, with $\Psi_t(q)$. For other wind directions $\Psi^*_c(q)$ lies between these boundary functions.

For other components we will need the value of the integral:
\begin{equation}
\Psi_2(q) = \frac{2}{\pi}\int_{-\pi}^{+\pi} \sin^2(\pi b q\cos\phi)\sin^{4}\phi {\rm\,d}\phi  = \frac{3}{4} - \frac{6J_2(2\pi b q)}{(2\pi b q)^2}.
\end{equation}
Using relation $\Psi^*_l(q) = \Psi^*_c(q) - \Psi^*_t(q)$, the following expressions for $l$- and $t$-motions can be obtained:
\begin{equation}
\Psi^*_l(q) = \frac{1}{2}\Psi_l(q) + \cos 2\theta\,\Bigl(\frac{1}{2}\Psi_l(q)-\Psi_t(q)+\Psi_2(q)\Bigr).
\label{eq:psil2}
\end{equation}
\begin{equation}
\Psi^*_t(q) = \frac{1}{2}\Psi_t(q) + \cos 2\theta\,\Bigl(\frac{1}{2}\Psi_t(q)-\Psi_2(q)\Bigr).
\label{eq:psit2}
\end{equation}

\section{Basic integrals}
\label{sec:integrals}

The expressions from Append.~\ref{sec:a1} reduce to tabular integral 3.715.21  \citep{grad_en}:
\begin{multline}
\mathcal I_n(z) = \int_0^\pi \cos(z\cos x)\,\sin^{2n} x {\rm\,d}x \\= \sqrt{\pi}\,\Gamma(n+1/2)(2/z)^n\,J_n(z).
\end{multline}
In calculations we used integrals with $n=0, 1, 2$ in limits $[-\pi$, $+\pi]$:  $\mathcal I_0(z) = 2\pi\,J_0(z)$, $\mathcal I_1(z) = 2\pi\,J_1(z)/z$ and $\mathcal I_2(z) = 6\pi\,J_2(z)/z^2$ in following form (as a rule):
\begin{multline}
\mathcal J_n(z) = \int_{-\pi}^{+\pi} \sin^2(z\cos x)\,\sin^{2n} x {\rm\,d}x \\ =
\frac{1}{2}\int_{-\pi}^{+\pi} (1-\cos(2z\cos x))\,\sin^{2n} x {\rm\,d}x.
\label{eq:Lint}
\end{multline}
In particular, $\mathcal J_0(z) = \pi(1-J_0(2z))$, $\mathcal J_1(z) = \pi(1/2-J_1(2z)/2z)$ and $\mathcal J_2(z) = \pi(3/8-3\,J_2(2z)/4z^2)$. Integrals with $\cos^{2n}(x)$ term obviously reduce to  $\mathcal J$-integrals. Integrals with $\sin$ to the odd power in mentioned limits are equal to zero.

\label{lastpage}
\end{document}